\documentclass[%
 reprint,
superscriptaddress,
nofootinbib,
 amsmath,amssymb,
 aps,
]{revtex4-2}

\usepackage{graphicx}
\usepackage{dcolumn}
\usepackage{bm}
\usepackage{braket}

\usepackage[]{todonotes}
\usepackage[toc,page]{appendix}
\usepackage{float}
\usepackage[hidelinks]{hyperref}

\begin{document}

\preprint{}



\title{AI-Assisted Discovery of Quantitative and Formal Models in Social Science} 

\author{Julia Balla}
    \email{jballa@mit.edu}
\affiliation{Department of Mathematics, Massachusetts Institute of Technology, Cambridge, MA 02139, USA}
\affiliation{AI Institute for Artificial Intelligence and Fundamental Interactions (IAIFI)}
\author{Sihao Huang}%
\affiliation{Department of Physics, Massachusetts Institute of Technology, Cambridge, MA 02139, USA}
\affiliation{St. Catherine's College, University of Oxford, Oxford, OX1 3UJ, UK}
\author{Owen Dugan}
\affiliation{AI Institute for Artificial Intelligence and Fundamental Interactions (IAIFI)}
\affiliation{Department of Physics, Massachusetts Institute of Technology, Cambridge, MA 02139, USA}
\author{Rumen Dangovski}
\affiliation{AI Institute for Artificial Intelligence and Fundamental Interactions (IAIFI)}
\affiliation{Department of Electrical Engineering and Computer Science, Massachusetts Institute of Technology, Cambridge, MA 02139, USA}
\author{Marin Solja\v{c}i\'{c}}
\affiliation{AI Institute for Artificial Intelligence and Fundamental Interactions (IAIFI)}
\affiliation{Department of Physics, Massachusetts Institute of Technology, Cambridge, MA 02139, USA}
   
\date{16 August 2023}
\begin{abstract}

In social science, formal and quantitative models, such as ones describing economic growth and collective action, are used to formulate mechanistic explanations, provide predictions, and uncover questions about observed phenomena. Here, we demonstrate the use of a machine learning system to aid the discovery of symbolic models that capture nonlinear and dynamical relationships in social science datasets. By extending neuro-symbolic methods to find compact functions and differential equations in noisy and longitudinal data, we show that our system can be used to discover interpretable models from real-world data in economics and sociology. Augmenting existing workflows with symbolic regression can help uncover novel relationships and explore counterfactual models during the scientific process. We propose that this AI-assisted framework can bridge parametric and non-parametric models commonly employed in social science research by systematically exploring the space of nonlinear models and enabling fine-grained control over expressivity and interpretability.
\end{abstract}

\maketitle

\label{sec:introduction}

Quantitative and formal models are widely employed in economics, sociology, and political science \cite{oliver_formal_1993, gailmard_formal_2012} for studying diverse subjects ranging from economic growth \cite{solow_contribution_1956} to participation in social movements \cite{granovetter}. They are also used to describe empirical regularities, such as patterns in voting behavior and city scaling laws, from which mechanisms can be derived and subsequently tested \cite{chatterjee_universality_2013, gabaix_zipfs_1999}. While qualitative methods remain a dominant paradigm in much of social science due to their ability to include normative arguments, capture greater complexity and nuance, and extrapolate from sparse data, mathematical models provide a complementary way to reason about mechanisms, generate questions, and produce empirically testable predictions \cite{epstein_why_2008}. 

These models are typically constructed using a combination of existing theoretical paradigms, intuition about the relevant variables, and insights on how they interact. Although intuition can serve as a good inductive prior, especially since researchers -- being human agents themselves -- can have valuable insights on social phenomena, it suffers from two key drawbacks. First, because available data can be high-dimensional, noisy, or strongly nonlinear, it may be difficult to recognize certain empirical regularities. Second, social phenomena are often explained by a large space of plausible and intuitive theories \cite{keohane_designing_2021}. This means that personal and experiential factors \cite{hammersley_bias_1997}, together with hidden assumptions made during the modeling process, can bias the models. The inclusion or omission of certain variables may even change the sign of the estimated parameters. These biases can threaten the external validity of these models, especially if they are tested in isolation without a systematic examination of the entire solution space \cite{healy_fuck_2017}.

Machine learning provides a collection of powerful tools for detecting patterns in data. However, these models are often uninterpretable and lack the ability to generalize beyond the observed data. Symbolic regression -- the process of discovering symbolic expressions from empirical data -- has thus evolved as a central method for uncovering hidden relationships in complex data. Most symbolic regression methods employ genetic algorithms, which perform a targeted search over populations of candidate solutions using strategies inspired by biological evolution~\cite{schmidt_distilling_2009, Keren2023}. Other successful approaches include sparse regression \cite{brunton_discovering_2016}, Bayesian learning \cite{guimera2020}, deep learning \cite{petersen2021deep, BiggioBNLP21}, or a combination of these methods \cite{udrescu_ai_2020, costa_fast_2021}. 
In recent years, they have seen substantive use in the discovery of fundamental laws in the natural sciences including mathematical conjectures \cite{keohane_designing_2021}, physical laws \cite{udrescu_ai_2020, cranmer_discovering_2020, wang_symbolic_2019}, and ecological models \cite{cardoso_automated_2020, martin_reverse-engineering_2018}.

\begin{figure*}
\includegraphics[width=1\textwidth]{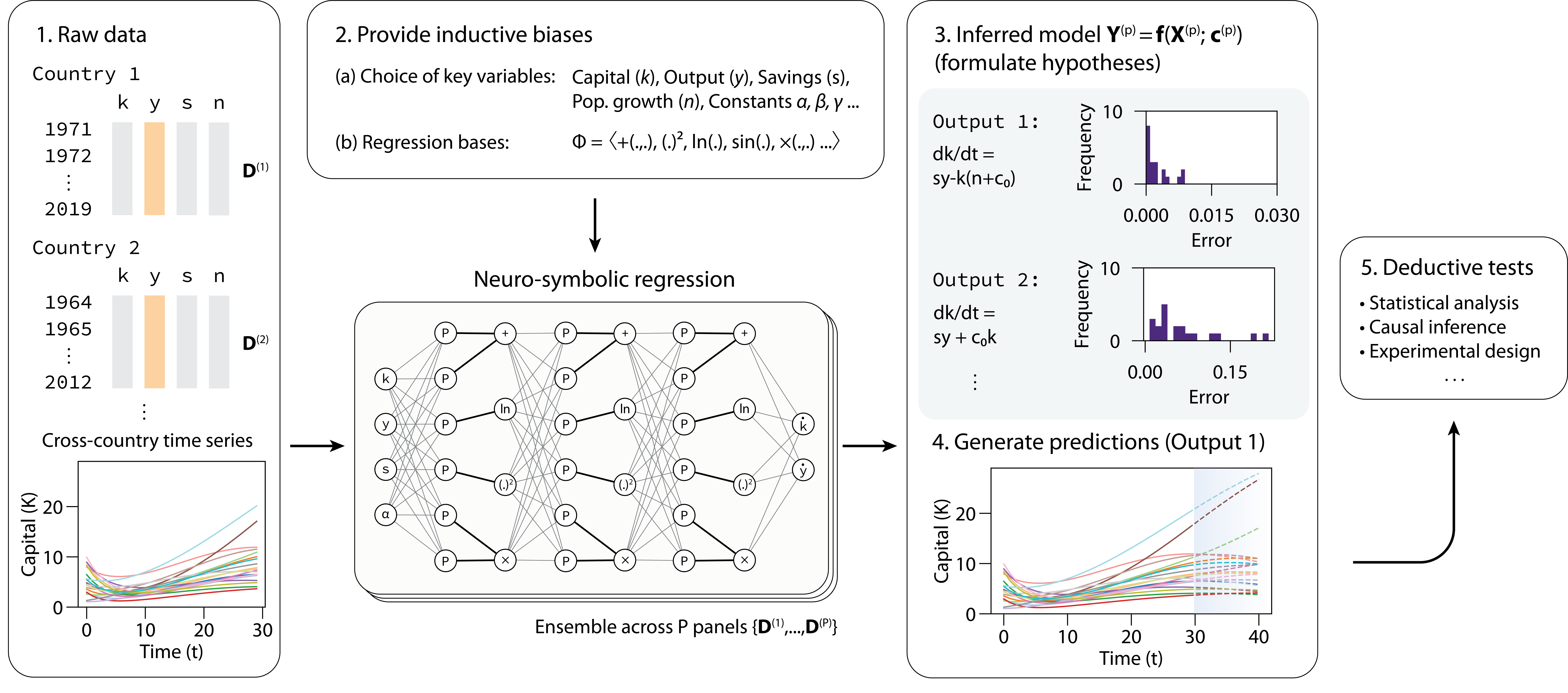}
\caption{\label{fig:workflow} Using neuro-symbolic regression to systematically guide model discovery in social science. Analogous to the inductive-deductive reasoning process, a dataset of interest (1) -- which may be time-series, cross-sectional, or longitudinal -- is supplied to OccamNet. The user can provide inductive priors (2), such as the choice of key variables, known constants, or specific functional forms to constrain the search space. OccamNet finds interpretable and compact solutions that model the input data by sampling functions from an internal probability distribution represented using \textit{P-nodes} \cite{costa_fast_2021}. In this example, OccamNet recovers the governing equation of the Solow-Swan model of economic growth \cite{solow_contribution_1956} from a synthetic dataset. Each formal model is characterized by its error distribution in the training set (3), allowing the user to identify outliers and interrogate its internal validity. The symbolic model is then used to generate predictions (4) to perform deductive tests across unseen data, either by partitioning a test set or informing experimental designs (5).}
\end{figure*}

There have also been various applications of symbolic regression to social science data, including the exploration of factor importance in carbon emissions \cite{pan_influential_2019} and unemployment rates \cite{riolo_explaining_2014}, modeling of oil production \cite{yang_modeling_2015}, and inference of strategic rules in repeated games \cite{duffy_using_2002}. Furthermore, techniques for fitting nonlinear dynamical systems models to data have been developed for studying complex phenomena such as school segregation \cite{spaiser_identifying_2018} or the relationship between democracy and economic growth \cite{ranganathan_bayesian_2014, ranganathan_understanding_2015}. However, most existing symbolic regression methods lack flexibility in incorporating user-specified inductive priors: a crucial element for seamless human-machine collaboration, particularly in cases where inference from sparse data can benefit greatly from human input. They are also unable to take advantage of the additional statistical power offered by longitudinal (panel) datasets, where information shared between panels can improve the generalizability of the learned model.

We address these problems in our work and propose a general framework for using neuro-symbolic regression to assist the discovery of quantitative models in social science. Building off \textit{OccamNet}~\cite{costa_fast_2021}, a neural-evolutionary method for efficient symbolic regression, our system searches for parsimonious functional relationships or differential equations and evaluates them according to their explanatory power. We select OccamNet for the symbolic regression backbone as it combines the advantages of genetic algorithms and deep learning-based methods and specializes in fitting unordered and small datasets, which are prevalent in the social sciences. Furthermore, unlike other neuro-symbolic regression methods, OccamNet is fast and scalable as it is able to seamlessly search expressions in parallel using modern AI accelerators, such as GPUs and TPUs, and to enable weight-sharing for regressing on several models at once. We thus leverage OccamNet to extend existing symbolic regression methods on two fronts: (i) improving noise tolerance in small datasets (ii) and enabling the system to learn symbolic relationships across longitudinal data (e.g., in a panel dataset across countries) using a weight-sharing approach. This allows one neural-symbolic model to be trained on multiple panels with different parameters simultaneously. Using the proposed workflow shown in Fig. \ref{fig:workflow}, these contributions allow neuro-symbolic methods to be more readily applied to \textit{ab initio} model discovery in social science, where small and noisy datasets are often available across contexts with different model parameters. 

We highlight several key applications of this system in a social science context. In section \ref{subsec:static}, we show that complexity regularization can be used to find higher-order corrections in network degree distributions, which we extend in mapping the complexity-accuracy Pareto front of economic production functions. We argue that such maps can help researchers find a trade-off among various quantitative models and debias the search process. Next, we apply this method to dynamical models in section \ref{subsec:dynamical}, where we use real-world, noisy data to discover the Lotka-Volterra equations and show that weight-sharing learning can be applied across epidemic waves to provide better performance in finding infection models. Finally, we highlight the utility of neuro-symbolic regression in evaluating existing models by using weight-sharing learning to map the complexity-accuracy front of economic growth models across 18 OECD countries and evaluate the generalizability of Solow and Swan's exogenous growth model \cite{solow_contribution_1956, swan_economic_1956}. We hope that the capabilities demonstrated in this article will inspire greater collaborative use of these machine learning methods in social science.

\section{Method}
\label{sec:occamnet}

Fig. \ref{fig:workflow} outlines our workflow for AI-assisted model discovery in social science. Symbolic regression allows us to uncover interpretable expressions as opposed to black-box or fully non-parametric results. These expressions, which may take the form of differential equations, scaling laws, or functional relationships, can shed light on the interactions between variables and help social scientists generalize beyond linear relationships. Furthermore, because symbolic laws can have system-specific constants, the inferred models also describe precise structural relationships that can be used to generate testable predictions across populations or timeframes. We employ a neural network architecture built on OccamNet to find sparse, interpretable formal models that fit the desired input data. In particular, we leverage OccamNet's architecture to implement two key features to systematically explore the space of candidate models in social science: complexity regularization and data ensemble learning.

In the most general formulation, OccamNet takes as input a panel dataset of the form $\mathcal{D} = \{\mathbf{D}^{(1)}, \ldots, \mathbf{D}^{(P)}\}$ on $K$ variables, where each dataset $\mathbf{D}^{(p)} \in \mathbb{R}^{N_p \times K}$ with $N_p$ samples is partitioned by the user into an input set
\begin{equation}
    \mathbf{X}^{(p)} = \begin{bmatrix}
    {\mathbf{x}_1^\mathrm{T}}^{(p)}\\
    \vdots\\
    {\mathbf{x}_{N_p}^\mathrm{T}}^{(p)}
\end{bmatrix} \in \mathbb{R}^{N_p \times K_x}
\end{equation}
and a target set 
\begin{equation}
    \mathbf{Y}^{(p)} = \begin{bmatrix}
    {\mathbf{y}_1^\mathrm{T}}^{(p)}\\
    \vdots\\
    {\mathbf{y}_{N_p}^\mathrm{T}}^{(p)}
\end{bmatrix} \in \mathbb{R}^{N_p \times K_y}
\end{equation}
for symbolic regression, i.e. $\mathbf{D}^{(p)} = \left[\mathbf{X}^{(p)} | \mathbf{Y}^{(p)}\right]$. When the data is not longitudinal, we simply have $p=1$ with one dataset $\mathbf{D} = \left[\mathbf{X} | \mathbf{Y}\right]$.

Alternatively, the user can also allow OccamNet to automatically find a partition by having it search through all possible combinations of inputs and targets and fitting an implicit function to the dataset(s). This mode of operation is useful if the user is seeking new empirical regularities in data and wishes to obtain a ranked list of candidate relations based on their error and compactness.

The goal of the regression problem is to find vectors of functions and hidden constant parameters $\mathbf{f}^{(p)} = [f_{1}(\cdot \ ; \mathbf{c}^{(p)}), \ldots, f_{K_y}(\cdot \ ; \mathbf{c}^{(p)})]$ that approximate or exactly specify each target variable $y_i^{(p)} = f_{i}^{(p)}\left(\mathbf{x}^{(p)}\right)$ with input vector $\mathbf{x}^{(p)} \in \mathbb{R}^{K_x}$ based on a predefined set of $B$ basis functions $\mathbf{\Phi}=\left\{\phi_{i}(\cdot)\right\}_{i=1}^{B}$ and maximum expression depth $d$ \footnote{Expression depth is defined as the maximum number of nested operations.}. These basis functions can have any number of arguments, be specified over different domains, or include constants to be optimized over (e.g., power laws of the form $x^c$). They may include pre-determined variables, such as $\pi$, $e$, the measured rate of capital growth, or the basic reproduction number of a virus. Furthermore, OccamNet can operate with piece-wise functions or non-differentiable, potentially recursive programs like $\text{MIN}(x_0, x_1, x_2 \ldots)$, enabling a wide range of possibilities for constructing formal models. 

OccamNet works by representing a probability distribution over a space of user-specified functions. During training, OccamNet samples $S$ functions from its probability distribution, evaluates their effectiveness at fitting the given data, and updates its probability distribution so as to weight its probability toward better-fitting solutions. An overview of the implementation of OccamNet and its advantages over other neuro-symbolic regression methods for social science data is provided in Appendix A.

\subsection{Regularization}
\label{sec:regularization}

Complexity regularization enables model discovery to follow the maxim that ``the theory should be just as complicated as all our evidence suggests" \cite{keohane_designing_2021}. This requirement does not necessarily depend on the assumption of parsimony: although it may be true that simpler theories are more probable \cite{zellner_basic_1984}, it also helps to ensure, independently, that there is sufficient evidence in relation to the complexity of the model to yield clear and specific research designs. Furthermore, in the computational context, the function relating variables of interest often only consists of a few terms. This allows us to frame symbolic regression as a sparse regression problem, making model discovery computationally tractable and avoiding the need to perform a brute-force search over functions \cite{brunton_discovering_2016}. 

To bias OccamNet toward simpler solutions, we include two regularization terms in the optimization objective, which the user can tune to control the simplicity of OccamNet's output: \textit{Activation Regularization} and \textit{Constant Regularization}. These terms penalize the number of activation bases and the number of tune-able constants used in the resulting expression respectively.

We further constrain the search space by exploiting dimensional analysis: a powerful technique in physics that is used to transform a problem into a simpler one with fewer variables that are all dimensionless. For instance, a dataset on economic growth might include variables with units such as dollars per capita or units of labor per year. Upon user specification of the units of each input and target variable, OccamNet checks whether the units of a sampled function match those of the target variable. It then adds a \textit{Unit Regularization} term to penalize functions where the units do not match. While OccamNet may also add a unit-normalizing constant in front of the mismatched terms, this would come at the cost of additional complexity. The implementation of all of the regularization terms is described in more detail in Appendix \ref{appendix:regularization}. 

\subsection{Data ensemble learning by weight-sharing}

Unlike other symbolic regression contexts, social science data often contain both cross-sectional and time series information. Take, for example, panel data containing economic indicators across several countries. Each country has a set of hidden constant parameters inaccessible to the researcher. Even if the variables of interest are related by the same functions, solutions to the dynamical system may be qualitatively very different because each country has different parameters and initial conditions. Learning on ensembles of data by sharing the neural network weights of OccamNet on multiple panels allows us to aggregate the statistical power of such datasets and reconstruct the dynamical equations even if there are hidden parameters, and if individual country trajectories do not traverse the full phase space of the system. 

Consider a panel dataset across $P$ panels, $\{\mathbf{X}^{(1)}, \ldots, \mathbf{X}^{(P)}\}$ where the rows of each panel are time-dependent vectors of variables ${\mathbf{x}}^{(p)}(t)$. Additionally, let each panel have a vector of hidden parameters $\mathbf{c}^{(p)} = [\alpha^{(p)}, \beta^{(p)}, \ldots ]^{\top}$ that is inaccessible to the researcher.
OccamNet is able to learn on ensembles of data by fixing a shared set of functions $\mathbf{f}(\cdot)$ for all panels and fitting the optimal constants $\mathbf{c}^{(p)}$ for each panel individually. The system repeats this step for a specified number of epochs and selects the optimal functions $\mathbf{f}^{*}(\cdot)$ with the lowest mean squared error with respect to the target data. To support ensembles with datasets of varying sizes, we normalize the total error by dividing each individual dataset error by the number of samples in the given panel.

Since the loss computed at each epoch is a distribution across the panels, different summary statistics may be used to select the optimal symbolic model. For instance, the loss may be computed as the mean or median. However, because the model $f(\mathbf{x})$ is generally nonlinear, errors of the form $f(\mathbf{x} + \boldsymbol{\epsilon})$, where $\boldsymbol{\epsilon}$ is a vector of Gaussian distributed entries with zero mean, can lead to highly skewed distributions that require alternate measures of central tendency. It may also be the case that certain outliers require additional variables or theoretical models to explain. In this case, the researcher can choose to identify the optimal formal model from the error histogram. The data ensemble learning modality is demonstrated in Fig. \ref{fig:workflow}.

\subsection{Differential equation discovery}

For time-dependent datasets, we wish to discover dynamical systems of the form $\dot{\mathbf{x}}(t)=\mathbf{f}(\mathbf{x}(t))$, where $\mathbf{x}(t)$ is a vector denoting the state of the system at time $t$. The input to OccamNet is a time-series dataset
\begin{equation}
    \mathbf{X} = \begin{bmatrix}
    \mathbf{x}^\mathrm{T}(t_0) \\
    \mathbf{x}^\mathrm{T}(t_0+h)\\
    \vdots\\
    \mathbf{x}^\mathrm{T}(t_0 + (N-1)h)\\
  \end{bmatrix} \in \mathbb{R}^{N \times K_x}
\end{equation}
where we have $N$ samples that are timestep $h$ apart starting from an initial state at $t_0$.
Our regression target is then given by $\mathbf{Y} = \dot{\mathbf{X}}$.

Fitting dynamical equations follows the same process as described in \ref{sec:occamnet} with the exception that $\mathbf{Y}$ is first computed numerically from the input data using the central difference formula\footnote{The central difference formula is given by $f'(x) \approx \left(f(x+h)-f(x-h)\right)/2h$ and accounts for second-order error in the derivative approximation.}. Since each $\mathbf{x}(t)$ is often contaminated with noise, it may be necessary to filter both $\mathbf{X}$ and $\dot{\mathbf{X}}$ to counter differentiation error, as described in Appendix \ref{appendix:diff_eq}.

While we primarily evaluate OccamNet on systems of non-linear differential equations in which the target variables are decoupled, we note that our model can also be applied to coupled systems by fitting the equations individually and adding the output of each fit as an input variable. For instance, for a system of two ODEs
$\frac{dx_1}{dt} = f_1(x_1, x_2)$ and $\frac{dx_2}{dt} = f_2(x_1, x_2, \frac{dx_1}{dt})$ as regression targets, one can use OccamNet to fit $\frac{dx_1}{dt}$ and append the predicted time-series to the input data for fitting $\frac{dx_2}{dt}$.

\subsection{Method scalability}

Let $S$ be the number of sampled functions at each layer, $N$ be the total number of points in the input dataset, $K_x$ and $K_y$ be the number of input and target variables in the dataset, $B$ be the maximum number of activation functions per layer, and $D$ be the maximum expression depth. Then the total computational complexity of an OccamNet network with the ability to express $B^D$ possible expressions is bounded as 
\begin{equation}\label{eq:complexity_bound}
    O\left(SN\left[DB + K_x \right]\left[DB + K_y \right]\right).
\end{equation}
The full derivation is provided in Appendix \ref{appendix:complexity}. This bound implies that, holding all other parameters fixed, the complexity scales linearly in the number of functions sampled as well as in the size of the dataset, allowing our method to be used for large social science tasks. While the complexity scales quadratically in the number of basis functions per layer and in the  depth, a crucial advantage of OccamNet is that it represents complete expressions with a single forward pass, thereby enabling sizeable gains in speed when using GPUs. A more detailed analysis of OccamNet's scalability is provided in \cite{costa_fast_2021}.




\begin{figure*}
\includegraphics[width=1\textwidth]{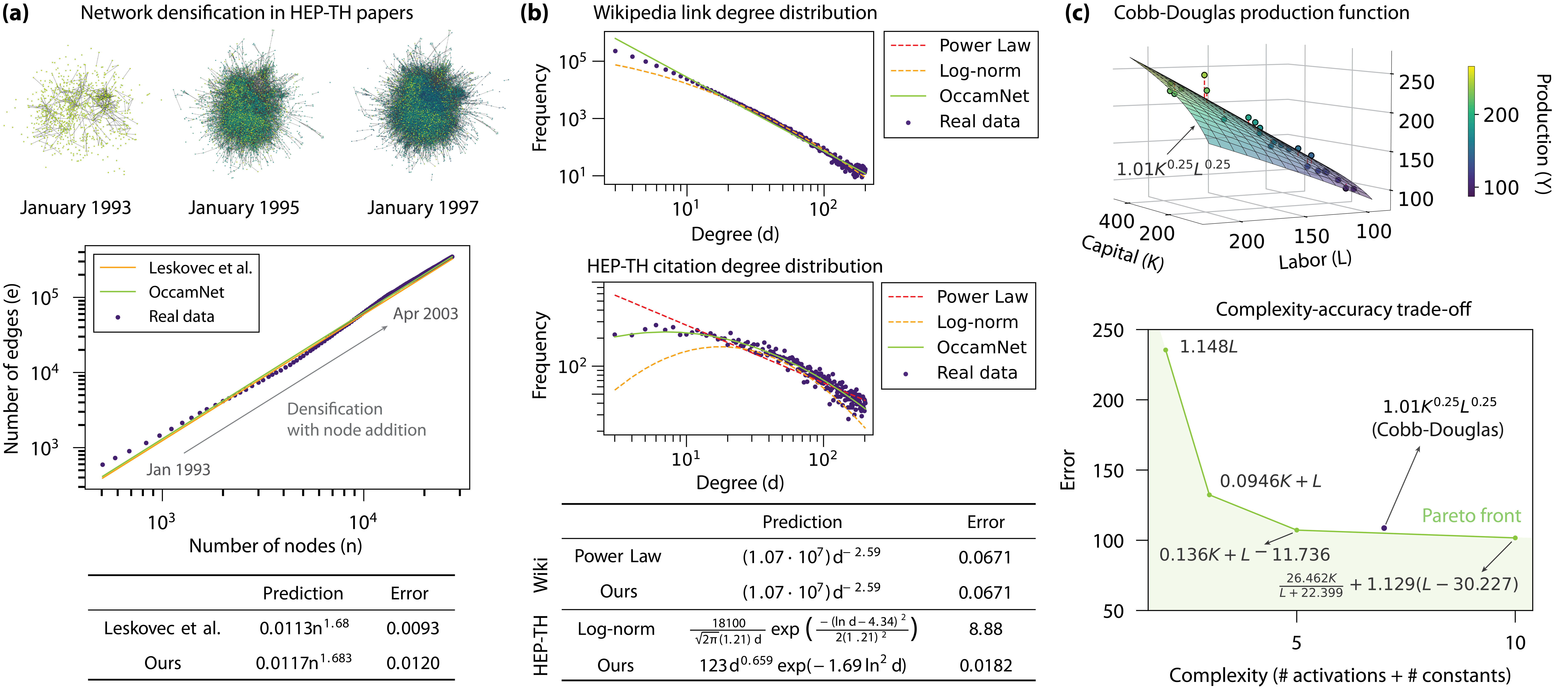}
\caption{\label{fig:static} \textbf{(a)} Discovering the network densification process in paper citations. The top diagrams show the evolution of the arXiv high energy physics citation network over time. OccamNet is able to discover the correct functional form for the densification law $a x^{\beta}$ based on a table of network properties \cite{leskovec_citations}. \textbf{(b)} A similar experiment is performed to discover the scaling law in the rank-size distribution of node degrees in the arXiv citation and Wikipedia hyperlink networks. At the same level of complexity regularization, OccamNet picks the simplest solution for each dataset, demonstrating the principle of Occam's razor. Our model finds a power law for the Wikipedia network and more complex function for the citation network.
\textbf{(c)}
We apply OccamNet to real-world, multi-dimensional economic data from Cobb and Douglas's 1928 paper. The complexity-accuracy trade-off is evaluated by sweeping across the available modes of regularization.
}
\end{figure*}

\section{Case studies} \label{subsec:studies}

We validate our AI-assisted approach on synthetic and real-world datasets related to social science covering both static distributions and dynamical systems. All dataset pre-processing details and model parameters are available in Appendix \ref{appendix:case_studies}.

\subsubsection{Static distributions}\label{subsec:static}

We first evaluate our neuro-symbolic regression method for the discovery of social science models on static distributions without time dependence. We extract symbolic relations as illustrated in Fig. \ref{fig:static}, which summarizes our results on fitting datasets of single variables such as network densification and degree distributions, as well as a multi-variable distribution for the Cobb-Douglas production function \cite{cobb_theory_1928}.\\

\noindent \textbf{Network properties.} Beginning with Barabási and Albert's seminal paper on scale-free networks \cite{barabasi_emergence_1999}, graph formation properties and degree distributions have become an increasingly active area of research. Network structures are studied in economics and sociology in the context of clustering, information spread, and skewed distributions, such as those exemplifying the ``rich get richer" effect in preferential attachment processes. Here, we focus on two sets of persistent properties observed in human networks: densification laws and degree distributions. 

In Fig.~\ref{fig:static} (a), we use neuro-symbolic regression to find the network densification process in citation networks. Densification occurs when the number of edges in a graph grows super-linearly in the number of nodes over time.  We use OccamNet to reconstruct the densification law of the citation graph of high energy physics preprints on arXiv from January 1993 to April 2003, where the number of edges grows according to $e(t) \propto n(t)^a$ as shown by Leskovec et al. \cite{leskovec_graphs_2005}. Our fit closely matches their baseline with a densification law exponent of approximately $1.692$, indicating that the number of citations in each paper on average grows over time. Note that we expect OccamNet to produce a fit with slightly higher error compared to the Ordinary Least Squares solution since our model has the additional task of finding the correct functional form before fitting the constants using gradient descent. For more accurate constant parameters, one may opt to take the functional form outputted by OccamNet and then fit the parameters using non-linear least squares.

Next, we use OccamNet to discover the functional form of various network degree distributions. Power law distributions of the form $k^{-\alpha}$ have been found in contexts ranging from hyperlink graphs \cite{albert_diameter_1999} to biological networks \cite{albert_scale-free_2005}. These distributions have been proposed to be universal across a wide range of systems, spawning extensive literature about their formation mechanisms and their links to network dynamics. However, alternative degree distributions like log-normal, Weibull, and exponential have also been observed in real-world networks, stimulating new work on finding alternative mechanisms that explain these functional forms \cite{broido_scale-free_2019}. Employing neuro-symbolic regression, we can go beyond the standard set of distributions to find any function drawn from the specified bases.

Fig.~\ref{fig:static} (b) shows the degree distribution for both Wikipedia article links \cite{wikipedia2009} and the high energy physics citation network. OccamNet is trained on both datasets using the same complexity regularization. In the Wikipedia dataset, the system discovers the power law distribution independently, while in the citation dataset, it finds that the degree distribution is better fit by a function of the form $k^\alpha e^{-c \ln^2(k)}$. The two examples underscore the principle of Occam's razor. For a given level of complexity regularization, the system picks the simpler functional form for the Wikipedia dataset because this dataset more closely resembles a power law. On the other hand, OccamNet discovers a more complex, but far more accurate expression for the HEP-TH dataset because this dataset deviates significantly from simpler models. The citation network fit performs better than both the power law and log-normal distributions.

It is worth noting that using mean-square error minimization does not generally provide the optimal estimator for any particular degree distribution. For instance, least-squares fitting can produce inaccurate estimates for power law distributions, and goodness-of-fit tests such as likelihood ratios and the Kolmogorov-Smirnov statistic must be used to evaluate the model \cite{clauset_power-law_2009}. However, in the case where the functional form is not known a priori, we opt to use mean-square error minimization as a Bayesian posterior with a non-informative prior. Other informative priors can be used when the set of functions are narrowed down, for instance, by performing maximum-likelihood estimation for each of the hypotheses proposed by the symbolic regression system. \\

\noindent \textbf{Production functions.}
\label{subsec:production} One of the best-known symbolic relations in economics is the Cobb-Douglas production function \cite{cobb_theory_1928}. Its introduction in 1928 represented the first time an aggregate production function was developed and statistically estimated \cite{biddle_retrospectives_2012}. Predicting that a stable relationship between inputs and outputs could be derived from empirical data, the authors found a robust linear relationship from the log output-to-capital ratio and the log labor-to-capital ratio of the American manufacturing industry. This relationship was later shown to hold across countries and timeframes \cite{douglas_cobb-douglas_1976}. The function has since been widely adopted as a measure of productivity, as a utility function, and as a foundation of subsequent models such as the Solow-Swan model of long-run economic growth \cite{solow_contribution_1956, swan_economic_1956}. 

In its most common form, the law states that the total production (output) $Y$ of a single good with two input factors, capital $K$ and labor $L$, is given by
\begin{equation}\label{eq:cobb-douglas}
    Y = AK^{\alpha}L^{\beta},
\end{equation}
where $A$ is typically a constant representing total factor productivity (ratio of output not explained by the quantity of inputs), and $\alpha$ and $\beta$ are the output elasticities of capital and labor, respectively. In cases where we expect to see constant returns to scale (i.e., an increase in the input quantity leads to an equivalent increase in the output), $\alpha$ and $\beta$
are additionally assumed to satisfy the constraint $\alpha + \beta = 1$.

Cobb and Douglas used a US manufacturing dataset covering annual production from 1899-1922 to fit the linear regression $\log(\frac{Y}{K}) = A
+ \beta \log(\frac{L}{K})$ by ordinary least squares. The resulting constant parameters were given by $A=1.01$ and $\beta=0.75$, resulting in the equation $Y=1.01K^{0.25}L^{0.75}$ which is plotted in Fig.~\ref{fig:static} (c). We explore the expressions for $Y$ predicted by OccamNet for the same dataset at varying levels of complexity regularization. We define the complexity value of an expression as the sum of the number of activation bases and constants. While there are other possible measures of complexity, such as the number of bits needed to store the symbol string of the expression \cite{udrescu_ai_2020}, our definition matches the regularization options provided for OccamNet training as described in \ref{sec:regularization}. We also note that any complexity metric will not be an exact measure of parsimony, which is typically defined according to the paradigmatic context through which the expression is interpreted. 

The Pareto front of the complexity-accuracy trade-off from sweeping across \textit{Activation Regularization} parameter values is shown in Fig. \ref{fig:static} (c). The full specification of model parameters is included in Appendix \ref{appendix:cobb}. Without any regularization, OccamNet discovers an expression with lower loss and higher complexity than the Cobb-Douglas baseline. Moreover, with a small amount of regularization, our model finds an equation with both lower loss and lower complexity than the baseline. The Cobb-Douglas function, therefore, lies outside the Pareto front -- under our particular definition of complexity -- for the given US manufacturing dataset. We recover a Cobb-Douglas-like expression of the form $K^{0.249}L^{0.754}$ only after restricting the set of possible bases for the regression to the ones found in \eqref{eq:cobb-douglas}. With only 24 available samples available in the data, our model benefits from a stronger inductive prior on the target function to reduce the risk of over-fitting. While there are strong theoretical reasons to prefer the Cobb-Douglas function, OccamNet offers a method for both discovering new models and for validating the existing expression.

This experiment reveals an important component of using AI-assisted model discovery in social science research. As discussed, the true definition of parsimony depends on how the expression is interpreted through an economic lens and not the length of the mathematical equation. How ``good" a model is depends on how it is used: in some contexts, a good theory is simple and explains the majority of the observations, while in others, it is highly complex and resolves all the empirical phenomena. 

Therefore, the use of symbolic regression must be viewed as a collaborative human-machine process. We can analogize model search as an optimization process across three dimensions: (i) the amount of variance explained by the model, (ii) its complexity, as per Occam's Razor, and (iii) the meaning behind the terms and structure of the model, as interpreted in the context of the field. While we cannot algorithmically search (iii), our method provides a way to fix a point along that dimension via a choice of inductive priors and execute a systematic search along the first two dimensions. Furthermore, restricting the search space by injecting existing knowledge about the model (performed by customizing the library of bases) can be beneficial to the regression tasks, particularly on sparse and noisy datasets. The key insight is that while humans are still needed to interpret the resulting models, computational methods offer a powerful way to systematize our search through two key dimensions in the hypothesis space.

\begin{figure*}
\includegraphics[width=1\textwidth]{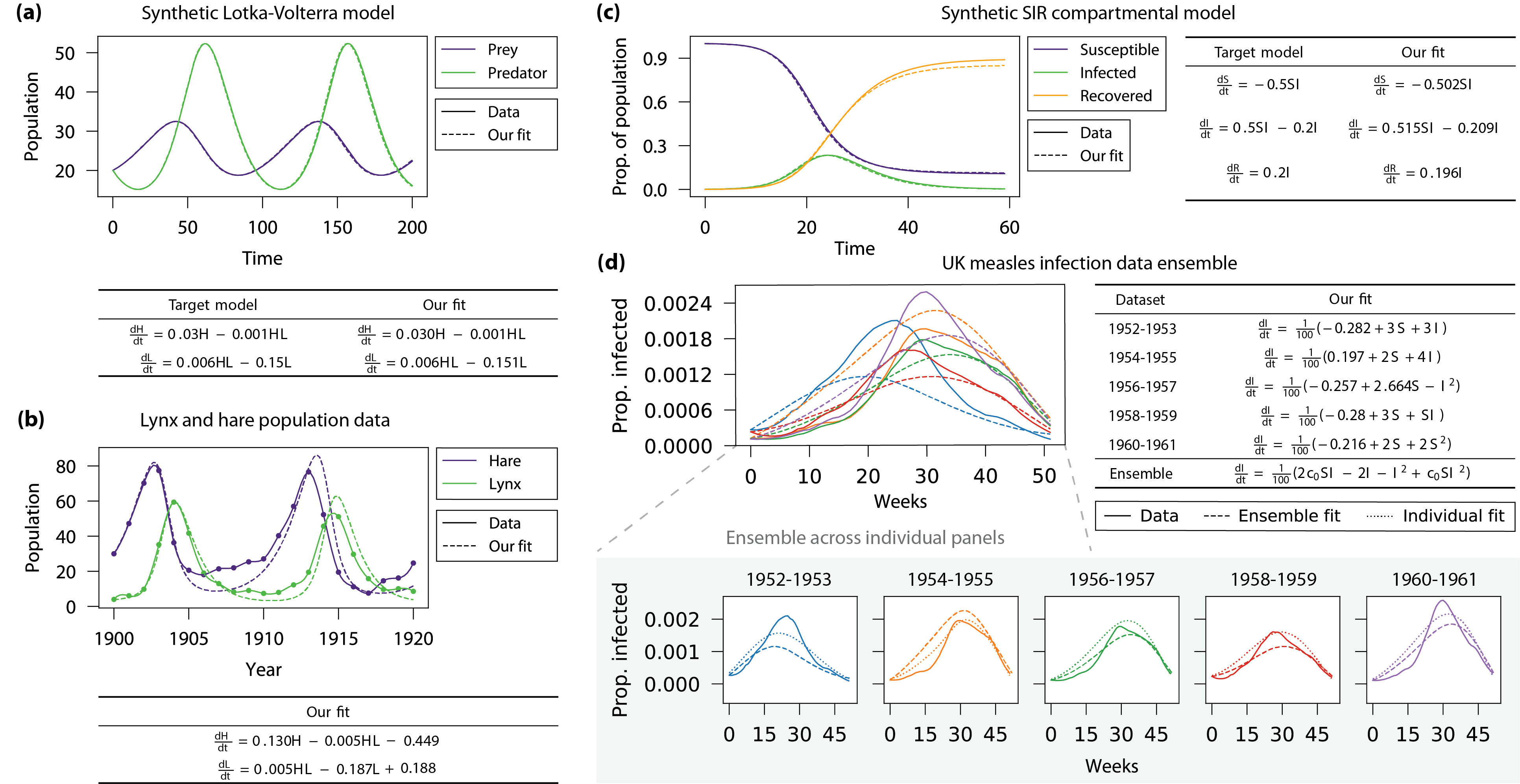}
\caption{\label{fig:dynamic} Regression of coupled dynamical models using noisy real-world data. \textbf{(a)} Time-series plot of a simulated Lotka-Volterra predator-prey system. OccamNet was able to correctly reconstruct the functional form and constants with high accuracy. \textbf{(b)} Using cubic spline interpolation, our system was able to learn the two differential equations from noisy, real-world data of lynx and hare populations with just 21 data points each. The inferred non-linear model can then be used to extend predictions of future populations. \textbf{(c)} The symbolic regression system is used to infer the SIR model of pandemic spread in synthetic data and \textbf{(d)} an ensemble of real-world measles infection data in the UK.}
\end{figure*}

\subsubsection{Dynamical laws}
\label{subsec:dynamical}

We next apply our neuro-symbolic regression method to differential equation discovery. OccamNet is primarily evaluated on systems of non-linear differential equations, where each ODE is fit independently. We use both synthetic and real-world datasets to fit models of population dynamics, disease spread, and economic growth.\\

\noindent \textbf{Predator-prey relationships.} Ecologists have long relied on mathematical models of population dynamics for studying ecological processes such as predator-prey interactions. While these models -- often taking the form of differential equations -- were historically derived from first principles and rigorous theory-building, symbolic regression offers a promising tool for reverse-engineering ecological models from data \cite{martin_reverse-engineering_2018}.

A simple yet paramount model of predator-prey interactions between two species is given by the Lotka-Volterra equations \cite{lotka_contribution_1910},
\begin{equation}
    \begin{aligned}
        \frac{dH}{dt} &= \alpha H - \beta H L \\
    \frac{dL}{dt} &= \delta H L - \gamma L,
    \end{aligned}\label{eq:lotka-volterra}
\end{equation}
where $H$ and $L$ are the populations of the prey and predator, respectively. The constant parameter $\alpha$ is the rate of exponential population growth for the prey, $\gamma$ is the rate of exponential death for the predator, $\beta$ is the rate at which the predators destroy the prey, and $\delta$ is the rate of predator population growth due to the consumption of prey. Inspired by the chemical law of mass action \cite{berryman_orgins_1992}, the Lotka-Volterra equations state that the rate of predation is directly proportional to the product of the populations of the prey and predator.

The Lotka-Volterra predator-prey model has significant implications beyond ecology. Richard Goodwin was one of the first to adopt the model to describe the cyclical relationship between employment and wages in economics, otherwise known as Goodwin's class struggle model \cite{gandolfo_lotka-volterra_2007}. Other applications of generalized Lotka-Volterra systems in economics include the modeling of macroeconomic indicators \cite{sterpu_generalization_2023}, as well as wealth distributions in society and values of firms in the stock market \cite{malcai_theoretical_2002}. 

In Fig. \ref{fig:dynamic} (a) we simulate a system of Lotka-Volterra equations as defined in \eqref{eq:lotka-volterra} with synthetic parameters $\alpha$, $\beta$, $\gamma$, and $\delta$. We then use OccamNet to fit $\frac{dH}{dt}$ and $\frac{dL}{dt}$ and simulate our generated differential equations using the same initial conditions. Our model is able to rediscover the exact functional form and constant parameters.

We then apply OccamNet to recover the Lotka-Volterra model from the Hudson Bay Company's data on the lynx and hare population in Canada from 1845-1935 \cite{maclulich_fluctuations_1937}. As shown in Fig. \ref{fig:dynamic} (b), we use the subset of records from 1900-1920 as it contains two population cycles with the least apparent noise. We additionally apply cubic-spline interpolation to reduce noise before fitting the numerically computed derivatives as discussed in Appendix \ref{appendix:diff_eq}.

When fitting the interpolated data using OccamNet, we recover a system of ODEs that closely resembles the Lotka-Volterra equations. In particular, our model is able to learn the exponential population growth and decay terms $\alpha H$ and $-\gamma L$, as well as the critical interaction terms of the form $-\beta HL$ and $\delta HL$. Additionally, the best OccamNet fit includes small constant correction terms of $-0.449$ and $0.188$. These constants have much smaller magnitudes relative to the other terms and may be due to over-fitting. A researcher may choose to either ignore these kinds of correction terms, or instead increase the level of constant regularization in OccamNet to obtain a fit with higher loss but potentially better generalizability.\\

\noindent \textbf{Epidemic spread.} Next, we use OccamNet to discover compartmental models of epidemics. In particular, we use synthetic and real datasets for the well-known Susceptible-Infected-Recovered (SIR) \cite{kermack1927contribution} model for fixed population sizes, which is given by the ODE system
\begin{equation}
    \begin{aligned}
        \frac{ds}{dt} = -\beta i, \quad \frac{di}{dt} = \beta is - \gamma i, \quad \frac{dr}{dt} = \gamma i,
    \end{aligned}\label{eq:sir}
\end{equation}
where $s(t)$, $i(t)$, and $r(t)$ represent the proportions of the susceptible, infected, and recovered populations at time $t$ respectively, and where $s + i + r = 1$. 

The SIR model and numerous variants are often used to describe the spread of infectious diseases such as measles \cite{bjornstad_dynamics_2002} or COVID-19 \cite{cooper_sir_2020}. Such models are valuable for predicting the magnitude or duration of an epidemic, estimating epidemiological parameters such as the basic reproduction number \cite{van_den_driessche_reproduction_2017} (given by $R_0 = \frac{\beta}{\gamma}$ in the SIR model), and for forecasting the impact of public health interventions. Beyond modeling disease spread, SIR variants have also been used to study the diffusion of information on social networks \cite{woo_epidemic_2016, kumar_information_2021}, and thus carry substantial relevance to social science.

In Fig. \ref{fig:dynamic} (c), we simulate a synthetic time-series using the SIR model. OccamNet discovers the correct functional form of the SIR model along with approximately correct constant parameters up to rounding. The deviations of the learned constant parameters likely stem from higher-order errors in the numerical derivative estimate which are not addressed by the central difference formula. One could opt to use higher-order derivative approximations to account for such errors if necessary.

Next, we demonstrate the data ensemble learning functionality of OccamNet on a panel dataset for measles infection data in the UK. Horrocks and Bauch used the SINDy (Sparse Identification of Nonlinear Dynamics) algorithm \cite{brunton_discovering_2016} to fit a variant of the SIR model with time-varying contact rates to this measles dataset as well as to chickenpox and rubella data \cite{horrocks_algorithmic_2020}. We note that SINDy requires the specification of a pre-set library of possible terms for which the algorithm learns sparse vectors of coefficients. OccamNet instead uses a specified library of simple \textit{bases} to compose more complex terms with no additional inductive prior necessary. Our method requires less prior knowledge about the expression structure and is thus better suited to deriving new functional forms.

Horrocks and Bauch fit the entire measles time-series to a singular equation for $i(t)$ with time-varying $\beta(t)$ \cite{horrocks_algorithmic_2020}. We instead demonstrate OccamNet's ability to discover the SIR model for each cycle of the epidemic with different $\beta$ and $\gamma$ parameters. Using a subset of the data from 1948-1958, we generate an ensemble of five measles incidence cycles. We then apply the denoising techniques as in \cite{brunton_discovering_2016} and fit each dataset both individually and as an ensemble in which we learn the same functional form for all five datasets with varying constant parameters.

Fig. \ref{fig:dynamic} (d) highlights the advantage of ensemble learning over individual fits. When each 2-year cycle is fit independently, OccamNet struggles to learn expressions with the SIR-like form of $\frac{di}{dt} = \beta is - \gamma i$. Due to the sparsity and noisiness of each individual dataset, it only extracts the interaction term $\beta is$ from one of the five periods. The ensemble fit, however, discovers a function that included the key form of $\beta is - \gamma i$, as shown in the last row of the table in Fig. \ref{fig:dynamic} (d). The parameter $c_0$ is a placeholder for a constant that varies for each cycle. Ensemble learning can therefore help avoid over-fitting to individual datasets and improve generalization. While our model also learns higher order terms such as $i^2$ and $c_0si^2$ for the ensemble, these terms are of much smaller magnitude compared to the leading terms and are thus of less importance to the correct fit. This is another example in which OccamNet's custom regularization capabilities could be applied to eliminate higher-order terms.\\

\noindent \textbf{Long-run economic growth.}
\label{subsec:solow} The final dynamical system we consider is the neoclassical Solow-Swan model of economic growth \cite{solow_contribution_1956}. The model postulates that long-run economic growth can be explained by capital accumulation, labor growth, and technological progress. The model typically assumes Cobb-Douglas-type aggregate production with constant returns to scale, given by
\begin{equation}\label{eq:solow_y}
    y(t) = k(t)^\alpha
\end{equation}
where $y(t)$ is the output per unit of effective labor, $k(t)$ is the capital intensity (capital stock per unit of effective labor), and $\alpha$ is the elasticity of output with respect to capital. The central differential equation in the Solow model describes the dynamics of $k(t)$,
\begin{equation}\label{eq:solow_k}
    \frac{dk}{dt} = sy - (n+g+\delta)k,
\end{equation}
where $s$ is the savings rate, $n$ is the population growth rate, $g$ is the technological growth rate, and $\delta$ is the capital stock depreciation rate. This equation states that the rate of change of capital stock is equal to the difference between the rate of investment and the rate of depreciation. The key result of the Solow growth model is that a greater amount of savings and investments do not affect the rate of economic growth in the long run.

While the Solow model was originally derived to describe U.S. economic growth, it has also been applied to other countries. If the model were perfectly universal, we would expect to see every country's capital grow according to equation \eqref{eq:solow_k}, with varying hidden parameters. We thus generate a synthetic example of country panel data for economic growth by simulating the Solow equations for $k(t)$ and $y(t)$, with the goal of rediscovering equation \eqref{eq:solow_k}. As an additional level of complexity, we model the savings rate $s$ and population growth rate $n$ as time-dependent variables that grow according to the equations $\frac{ds}{dt} = 0.05$ and $\frac{dn}{dt} = 0.05n$. Each panel dataset is generated by randomly sampling initial conditions and parameters $g$ and $\delta$ from uniform distributions of values outlined in appendix \ref{appendix:synth_solow}. The resulting cross-country time series is displayed in Fig. \ref{fig:workflow}. As demonstrated in the figure, OccamNet is able to recover the exact equation for $\frac{dk}{dt}$ (denoted as Output 1) with varying parameter $c_0$ for each of the 20 synthetic panels.

\begin{figure}
\includegraphics[width=0.5\textwidth]{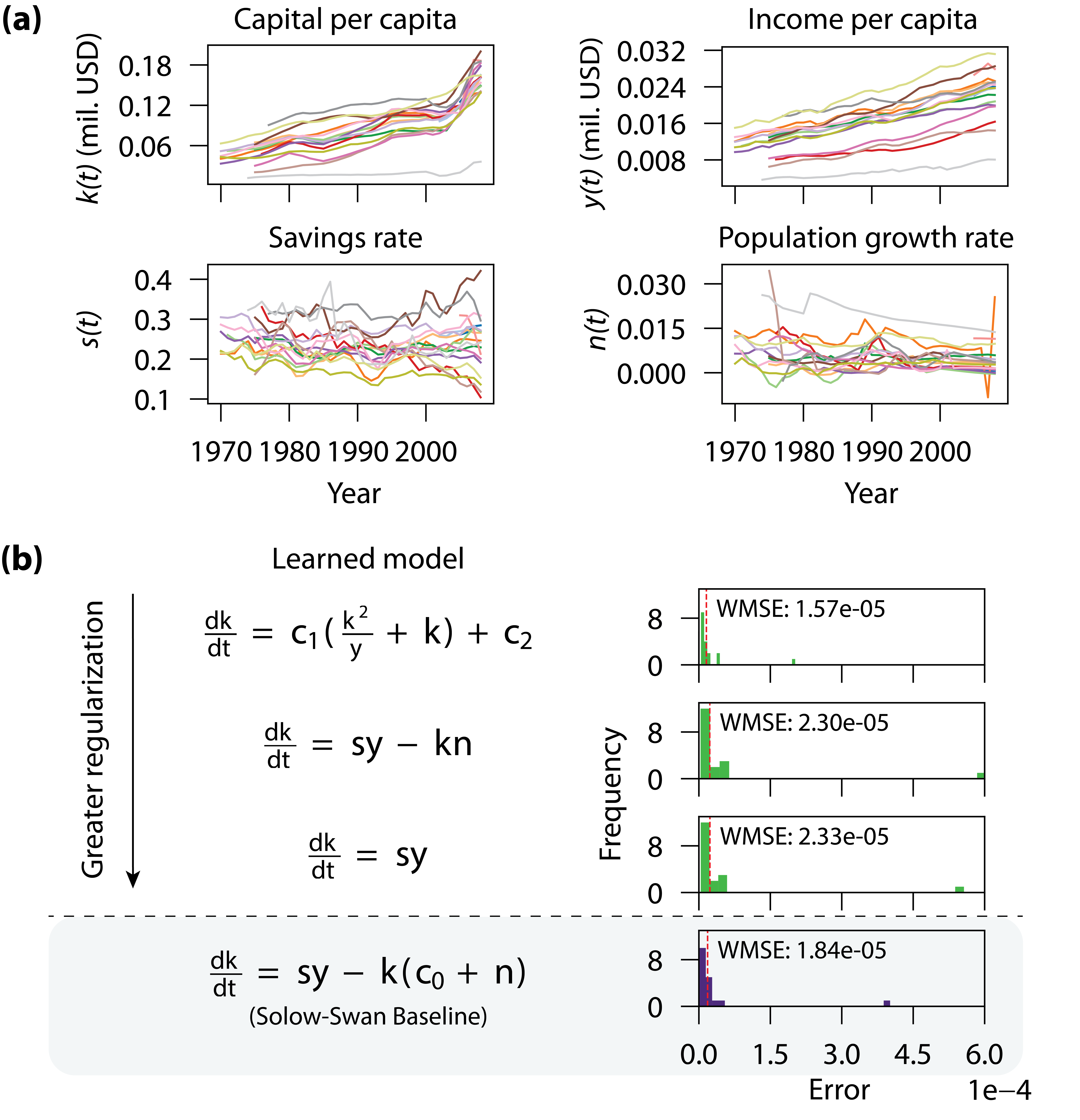}
\caption{\label{fig:ensemble} Ensemble learning of longitudinal (panel) macroeconomic data. 
\textbf{(a)} Country-level macroeconomic data on capital and income per capita, savings rates, and population growth for 18 OECD member countries. \textbf{(b)} Ensemble learning of the Solow economic growth model. The error distribution of the differential equation, applied to each country, is shown for three expressions generated with increasing levels of complexity regularization. The identification of outliers may inform alternative explanations, hidden parameters, or higher-order corrections to the economic model. }
\end{figure}

We then attempt to fit the Solow model of capital growth to real-world, noisy country data. The input to our symbolic regression includes data on capital per capita, income per capita, savings rate, and population growth compiled by Khoo et al. \cite{khoo_neural_2021}, where it was utilized for regression under the Green Solow model -- a variant of Solow-Swan for modeling sustainable growth \cite{brock_green_2004}. Following the methodology of \citep{brock_green_2004}, we select macro-economics data from 18 of the original 20 countries in the Organisation for Economic Co-operation and Development (OECD) for which data was available. The $k(t)$ data is originally sourced from the Penn World Tables \cite{feenstra_penn_table}, $y(t)$ and $n(t)$ are from the Maddison Project database \cite{maddison_2017}, and $s(t)$ is from the World Development Indicators \cite{world_bank}. There is no available data for the remaining parameters $g$ and $\delta$, so they are instead treated as learnable constants. We also apply Savitzky-Golay filtering to smooth the data before running the regression, as described in appendix \ref{appendix:real_solow}.

In Fig. \ref{fig:ensemble} (b), we compare the Solow-Swan model baseline to three expressions produced by OccamNet under increasing levels of complexity regularization. The Solow baseline is generated by finding the best-fit parameter $c_0=g+\delta$ in a least-squares of fit of equation \eqref{eq:solow_k}. The expression with no regularization is given by $\frac{dk}{dt}=c_1\left(\frac{k^2}{y}+k\right)-c_2$ with constants $c_1$ and $c_2$ that vary across countries. While this expression has a lower weighted mean-squared error (WMSE) than the baseline, its functional form does not carry immediate economic intuition like the Solow model. 

We then add \textit{Constant Regularization} as described in \ref{sec:regularization}, which results in the equation $\frac{dk}{dt}=sy-kn$ closely matching the functional form of \eqref{eq:solow_k}. This suggests that the Solow model has strong external validity as it can be discovered without any strong human priors. Finally, we apply \textit{Activation Regularization} in addition to \textit{Constant Regularization}, resulting in the output $\frac{dk}{dt}=sy$. The learned expression contains only one term from the Solow model and has the highest WMSE. The example in Fig. \ref{fig:ensemble} (b) concretely demonstrates the tradeoff between accuracy and simplicity in the discovery of symbolic models. A researcher would thus benefit from running OccamNet with several specifications of regularization to select a result with the desired level of precision and complexity.

\section{Discussion}

This article aims to test and demonstrate the usefulness of neuro-symbolic methods in social science. We argue that its contribution is twofold. First, human-engineered quantitative models remain a critical part of social science research as they provide a way to formalize intuition or qualitative arguments about a social system and bridge micro-macro theories. Recognizing that, neuro-symbolic methods help researchers benchmark these formal models against counterfactuals on the complexity-accuracy Pareto front: a practice not often done when new quantitative models are proposed. This is partly due to the difficulty of generating alternatives aside from baseline linear models or theories from the existing literature. The selection strategy of symbolic methods allows the intentional tuning of complexity (or meaningfulness, with human-aided post-selection) and prediction accuracy encountered by traditional parametric regression and deep learning methods.

Second, neuro-symbolic methods can also help social scientists discover new, interpretable models or generate novel hypotheses from data. Expressions on the Pareto-front can be interpreted and tested without prejudice. Aside from complete models, terms that repeatedly appear along the front may also uncover persistent interactions \cite{schmidt_distilling_2009}. Our proposed approach is powerful as it allows the user to tune the amount of bias entered into the model. While models developed \textit{ab initio} from a theoretical basis can be powerful in elucidating the implications of a present paradigm and incorporating human knowledge about human systems, the underconstrained nature of social science theories means that the researcher has to be conscious of their personal, experiential, or theoretical biases \cite{healy_fuck_2017, hammersley_bias_1997}. Using neuro-symbolic regression, the user can opt to inject a minimal amount of bias (e.g., provide a wide range of functional bases) or intentionally specify terms that existing theories may suggest.

Broadly speaking, two classes of quantitative models are typically employed in social science \cite{chen_revealing_2019}. Parametric regression has proven invaluable over the past few decades. Since data and computational resources have become widely available, researchers have made thorough use of methods like generalized linear models for inference \cite{nelder_generalized_1972}. Although parametric regression offers less expressive power than formal models, it provides benefits in interpretability and data efficiency, especially when there is strong reason to believe that the underlying relationship takes a simple (e.g., linear) form. Its key drawback, however, is that this interpretability and efficiency are derived from strong assumptions about how variables interact. More complex behaviors, such as temporal dynamics or nonlinear relations between multiple variables, are ubiquitous in social science but can be hidden by using simpler parametric methods. Likewise, nonlinear optimization methods for fitting specified models \cite{nelder_simplex_1965} also require an a priori structure.

As an alternative approach, social scientists can use non-parametric methods that make weak assumptions about the distribution of the data, such as kernel density estimation, local regression, or deep neural networks. These models are useful for prediction policy problems \cite{kleinberg_prediction_2015} -- where prediction accuracy, not causal identifiability, is prioritized -- and for capturing relationships in systems that often operate in highly nonlinear regimes. However, these methods come at the cost of interpretability. We suggest that neuro-symbolic regression can offer a bridge between these two classes of statistical methods. In between specifying a complete model to let the system search for optimal parameters and leaving it open-ended to seek a parsimonious expression, the user can provide any amount of human insight. For instance, they may specify the presence of a power-law relationship, an interaction term, or a derivative, and use the system to compose the rest of the model. In section \ref{subsec:studies}, we demonstrate each of these use cases and show that the framework can successfully help social scientists discover empirical regularities in the data with interpretable expressions.

Beyond interpretability, human-supplied biases are also useful in reducing the model search space, particularly when the problem faces a dearth of informative data. Since this is often the case in social science, we stress that the use of neuro-symbolic methods must be treated as a collaborative human-machine process in which a dictionary of motifs helps make inference problems tractable, as has been suggested by past work \cite{cardoso_automated_2020, multigene}. Humans also play a critical role in the validation and interpretation of the resultant models. Neuro-symbolic methods merely suggest functional fits; using these equations to paint a meaningful picture in the context of social science theory remains a human endeavor. The discovery of a formal model is often just the beginning, informing additional data collection, experiments, and qualitative inquiry \cite{epstein_why_2008}. A productive approach to quantitative social science needs to consider both the scientific and sociological role of modeling beyond its predictive capability alone. 

Neuro-symbolic regression holds the potential to be a powerful tool for helping social scientists with hypothesis creation and model discovery. An exciting next step would be to collaborate with social scientists in applying OccamNet to their research, particularly to datasets that are conjectured to capture a symbolic law. In problems with more available data and multiple independent panels, a machine learning approach allows data to be divided into training and validation sets -- provided that unexplained errors between the two sets are uncorrelated -- to improve the out-of-distribution performance, or external validity, of the learned model. We hope to see our proposed human-machine framework applied to unearth entirely new quantitative or formal models. Furthermore, OccamNet's flexibility in incorporating user-defined bases such as piecewise functions and logical operators allows for moving beyond learning arithmetic expressions. Future work could include applying OccamNet to learn more complex symbolic rules such as strategies in economic games \cite{duffy_using_2002} or network-structured models.

\section{Data availability}
The dataset for the HEP-TH citation network is publicly available online at \url{https://snap.stanford.edu/data/cit-HepTh.html} \cite{leskovec_citations, leskovec_graphs_2005}. The Wikipedia  hyperlink network data is available from the Network Repository at \url{https://networkrepository.com/web-wikipedia-link-en13-all.php} \cite{nr, wikipedia2009}. A compilation of the 1928 Cobb-Douglas dataset  \cite{cobb_theory_1928} is available at \url{http://web.mit.edu/pjdavis/www/options/cobbdouglas.xls}. MacLulich's 1937 compilation of the Hudson Bay Company's lynx-hare dataset \cite{maclulich_fluctuations_1937} is available online at \url{http://people.whitman.edu/~hundledr/courses/M250F03/M250.html}. The UK measles infection data is available at \url{https://math.mcmaster.ca/~bolker/measdata.html}. The compiled dataset for the Solow-Swan model experiments is available at \url{https://github.com/zykhoo/nODE-Solow} \cite{khoo2021neural}.

\section{Code availability}
The code used for our model and experiments is available at \url{https://github.com/druidowm/OccamNet_SocialSci}.

\begin{acknowledgments}
    We would like to thank Megan Yamoah for useful discussions during the drafting of this article. This work was sponsored in part by the United States Air Force Research Laboratory and the United States Air Force Artificial Intelligence Accelerator and was accomplished under Cooperative Agreement Number FA8750-19-2-1000. The views and conclusions contained in this document are those of the authors and should not be interpreted as representing the official policies, either expressed or implied, of the United States Air Force or the U.S. Government. The U.S. Government is authorized to reproduce and distribute reprints for Government purposes notwithstanding any copyright notation herein. This work was also sponsored in part by the National Science Foundation under Cooperative Agreement PHY-2019786 (The NSF AI Institute for Artificial Intelligence and Fundamental Interactions, http://iaifi.org/) and in part by the Air Force Office of Scientific Research under the award number FA9550-21-1-0317.
\end{acknowledgments}

\bibliography{references}

\onecolumngrid
\newpage
\appendix

\section{OccamNet}\label{appendix:occamnet}
OccamNet works by representing a probability distribution over a space of user-specified basis functions. During training, OccamNet samples functions from its probability distribution, evaluates their effectiveness at fitting the given data, and updates its probability distribution so as to weight its probability toward better-fitting solutions. Specifically, OccamNet consists of alternating \textit{Activation} and \textit{T-Softmax} Layers which together represent a probability distribution over a set of functions. The Activation Layers consist of user-specified basis functions, the building blocks out of which OccamNet constructs symbolic expressions. The T-Softmax Layers represent the probabilities of connecting an Activation Layer's outputs to the next layer, which is either another Activation Layer or the output layer. To generate trial functions, OccamNet samples from the T-Softmax Layer probabilities, obtaining a set of connections that represent a trial function: the input variables are traced along the generated graph and transformed by the basis functions along the path. Because the user specifies the bases included in each Activation Layer as well as other properties of the network such as the number of layers, OccamNet allows for a highly flexible user-specified search space. An example set of specified basis functions may be given by
\begin{equation}
    \mathbf{\Phi} = \{x+y, x-y, x\times y, x\div y, x^2, x^3, \sqrt{x}, x+c, x \cdot c, x^c, c_0\cdot x^2+c_1\cdot x+c_2, \log x, \exp x, \sin x, \cos x, \ldots\} 
\end{equation}
where $x$, $y$ are either variables or subexpressions and where $c$ is a placeholder for a constant parameter that is tuned via gradient descent. The user may also define custom bases that are relevant to the data. 

Additionally, the user may specify the maximum depth of the symbolic expression by stacking layers of $\mathbf{\Phi}$. The depth of an expression is given by the maximum number of nested basis functions. For instance, $\mathbf{\Phi} \times 2 $ indicates a maximum expression depth of 2, which allows for expressions such as $(x-y) + c$ but not $((x-y) + c) \times x$ (having a depth of 3). Moreover, each basis layer $\mathbf{\Phi}$ can contain a unique set of bases if desired.\\

OccamNet's architecture leads to a number of advantages over other symbolic regression methods: \begin{itemize}
    \item OccamNet allows the user to take full advantage of existing knowledge by encoding symbolic priors into the network. For example, if the user expects the expression to be a sum of distinct terms, they can set the final activation layer to include only addition bases. In many Symbolic Regression architectures, such inductive biases cannot be directly incorporated into the architecture.
    
    \item OccamNet's training procedure allows for a  non-differentable fitness term, thereby allowing the user to include discrete factors in the fitness, such as whether the units match or the number of constants used. Many neural approaches to Symbolic Regression cannot handle such factors.
    
    \item OccamNet is highly efficient and scales on a GPU, something that most symbolic regression methods cannot offer. This allows the user to find better fits in the same time as many other methods.
    
    \item OccamNet is a neural network that can be made fully differentiable, thereby enabling it to be combined with other neural architectures. Although we do not test this in our paper, OccamNet could, for example, be used to synthesize programs or to perform regression with image or text data. Thus, OccamNet can handle a broader range of potential Symbolic Regression tasks than many other approaches.
 
\end{itemize}

\subsection{Regularization terms}\label{appendix:regularization}
OccamNet uses two terms for complexity regularization: \textit{Activation Regularization} and \textit{Constant Regularization}. Let $\alpha[f]$ denote the number of activation functions $f$ and let $\gamma[f]$ denote the number of constants. The complexity regularization terms are thus given by
\begin{equation}
    w_\alpha \cdot \alpha[f] + w_\gamma \cdot  \gamma[f]
\end{equation}
where $w_\alpha$ and $w_\gamma$ are scalar weights for the respective regularization terms. OccamNet subtracts these terms from the unregularized fitness of a given expression $f$. The weight parameters  $w_\alpha$ and $w_\gamma$ allow the user to effectively customize the complexity of the resulting expression.

OccamNet also uses \textit{Unit Regularization} when indicated. For a dataset $\mathcal{D}$ with $K$ variables, the user may specify an array of exponents $\mathbf{U}\in \mathbb{R}^{K \times U}$, where $U$ is the number of unique units across the dataset, such that each entry $\mathbf{U}_{ij}$ contains the exponent of a unit $j$ for variable $i$. For example, if the $i$th variable has units of dollars/year, the $i$th row in $\mathbf{U}$ could be represented by the vector $[1, -1]$ where the first entry corresponds to dollars and the second entry corresponds to years. 

OccamNet then feeds these units through the sampled function. Each basis function receives a set of variables with units, may have requirements on those units for them to be consistent, and returns a new set of units. For example, $\sin(x)$ receives one variable which it requires to have units $[0,\ldots,0]$ and returns the units $[0,\ldots,0]$. Similarly, $+(\cdot,\cdot)$ takes two variables with units that it requires to be equal and returns the same units. Using these rules, OccamNet propagates units through the function until it obtains units for the output. If at any point the input units for a basis function do not meet that basis's requirements, that basis returns $[\infty,\ldots,\infty].$ Any basis functions that receive $[\infty,\ldots,\infty]$ also return $[\infty,\ldots,\infty]$. Finally, if the output units of $f$ do not match the desired output units (including if $f$ outputs $[\infty,\ldots,\infty]$), OccamNet marks $f$ as not preserving units.

The multiplication by a constant basis function, $x \cdot c$, can produce any output units. As such, it returns $[\text{NaN},\ldots,\text{NaN}].$ If any basis function receives $[\text{NaN},\ldots,\text{NaN}],$ it will either return $[\text{NaN},\ldots,\text{NaN}]$ if it has no constraints on the input units, or it will treat the $[\text{NaN},\ldots,\text{NaN}]$ as being the units required to meet the basis function's consistency conditions. For example, if the $\sin(\cdot)$ function receives $[\text{NaN},\ldots,\text{NaN}]$, it will treat the input as $[0,\ldots,0],$ and if the $+(\cdot,\cdot)$ function receives $[1,2,3]$ and $[\text{NaN},\text{NaN},\text{NaN}],$ it will return $[1,2,3]$. 

After sampling functions from OccamNet, we determine which functions do not preserve units. Because we wish to avoid these functions entirely, we bypass evaluating their normal fitness (thereby saving compute time) and instead assign a fitness of $-w_{\text{units}},$ where $w_{\text{units}}$ is a user-specified hyperparameter.

\subsection{Differential equation fitting}\label{appendix:diff_eq}
The target variables for differential equation fitting are computed numerically from time-series data. In particular, we use the central difference formula $f'(x) \approx \left(f(x+h)-f(x-h)\right)/2h$ which accounts for second-order error in the derivative approximation. We use this formula to compute the centered derivative of each variable (column) in the input data $\mathbf{X}$ to get $\dot{\mathbf{X}}$. 

When the time-series data is sufficiently noisy, it is helpful to counter differentiation error by applying pre-processing techniques such as total variance regularization \cite{rudin_nonlinear_1992} or the hard thresholding of singular values \cite{gavish_optimal_2014} as discussed in reference \cite{brunton_discovering_2016}. Alternatively, $\mathbf{X}$ may be filtered or regularized via frequency-based methods or polynomial interpolation prior to numerical differentiation. 


\subsection{Computational complexity}\label{appendix:complexity}

To derive the computational complexity of OccamNet, we first compute the complexity for each layer (i.e., a T-softmax and Activation layer block). Let $S$ be the number of sampled functions at each layer, $N$ the total number of points in the input dataset, $K_x, K_y$ be the number of input and target variables in the dataset, $B$ be the maximum number of activation functions per layer, and $D$ be the maximum expression depth. Let $L_{\text{img}}$ and $L_{\text{arg}}$ be the number of inputs and outputs of the $i$th T-softmax layer, respectively. Applying the sampled functions to the data has complexity $O(SNL_{\text{arg}})$, sampling the functions has complexity $O(SNL_{\text{arg}}L_{\text{img}})$ due to the computation of a softmax, and finally evaluating the probabilities also has complexity $O(SNL_{\text{arg}}L_{\text{img}})$. Therefore, the complexity of each layer is $O(SNL_{\text{arg}}L_{\text{img}})$.

For simplicity, we consider the case where the user specifies a fixed set of activation functions for each layer such that $L_{\text{arg}} = O(B)$ and $L_{\text{img}} = B\cdot i + K_x \leq BD +  K_x$. We may then write the total computational complexity as the sum of the complexities of $O(D)$ layers plus the complexity of the final layer,
\begin{align*}
    O(DSNB(DB+K_x) + SNK_y(DB+K_x)) &= O(SN\left[DB(DB+K_x) + K_y(DB+K_x) \right]) \\
    &= O\left(SN\left[D^2B^2 + DBK_x + DBK_y + K_xK_y\right]\right)\\ 
    &= O\left(SN\left[DB + K_x \right]\left[DB + K_y \right]\right).
\end{align*}

\section{Case studies}\label{appendix:case_studies}

Unless otherwise specified, all of the experiments are run with the following hyperparameters. 

\begin{table}[h]
\caption {Default OccamNet hyperparameters} \label{tab:defaul} 
\begin{center}
\begin{tabular}{>{\centering}p{0.1\textwidth}>{\centering}p{0.1\textwidth}>{\centering}p{0.15\textwidth}>{\centering}p{0.1\textwidth}>{\centering}p{0.18\textwidth}>{\centering\arraybackslash}p{0.18\textwidth}} 
 \hline
 Epochs & Learning rate & Constant learning rate & Decay rate & Activation regularization weight & Constant regularization weight\\ \hline
 1000 & 5 & 0.05 & 1 & 0 & 0\\
 \hline
\end{tabular}
\end{center}
\end{table}

For each experiment, we perform a grid-search of values for the remaining hyperparameters, which are described in the original OccamNet paper \cite{costa_fast_2021}. We search over the following parameter values and select the results with the lowest error:

\begin{itemize}
    \item Standard deviation ($\sigma$): [0.5, 5, 50]
    \item Top number: [1, 5, 10]
    \item Equalization: [0, 1, 5]
\end{itemize}

\subsection{HEP-TH citation densification}\label{appendix:densification}
We reconstruct the densification law of the citation graph of high energy physics preprints on arXiv  (HEP-TH) from January 1993 to April 2003, with 29555 papers and 352807 edges \cite{gehrke_ginsparg_kleinberg_2003}. For each month $m$ starting from January 1993, we aggregate the number of edges formed up to the end of $m$. We then transform the data as $\mathbf{x} = \log n(t)$ and $\mathbf{y} = \log e(t)$ to replicate the methodology of Leskovec et al. \cite{leskovec_graphs_2005} in fitting a line on the log-log scale. 

\begin{table}[H]
\caption {OccamNet hyperparameters for HEP-TH densification} \label{tab:densification}
\begin{center}
\begin{tabular}{>{\centering}p{0.25\textwidth}>{\centering}p{0.1\textwidth}>{\centering}p{0.1\textwidth}>{\centering\arraybackslash}p{0.1\textwidth}} 
 \hline
 Output &  Standard deviation & Top number & Equalization \\ \hline
 $e = 0.0117n^{1.683}$  & 5 & 5 & 5 \\
 \hline
\end{tabular}
\end{center}
\end{table}

Basis library: $\mathbf{\Phi} = \{+, -, \times, \div, x+c, x \cdot c\}\times 2$

\subsection{Degree distributions}\label{appendix:degrees}

\subsubsection{HEP-TH degree distribution}\label{appendix:hep-th_degree}

Using the same dataset as above, we fit the degree distribution of the citation graph at the last available time period. To mitigate the effects of the long tail of nodes with high degrees, we only include nodes with degree less than 200. We additionally remove the data points for degree 0 (papers with no citations) and degree 1. Finally, we fit the distribution in log-log space.

\begin{table}[H]
\caption {OccamNet hyperparameters for HEP-TH degree distribution} \label{tab:citations}
\begin{center}
\begin{tabular}{>{\centering}p{0.2\textwidth}>{\centering}p{0.1\textwidth}>{\centering}p{0.15\textwidth}>{\centering}p{0.1\textwidth}>{\centering}p{0.18\textwidth}>{\centering\arraybackslash}p{0.18\textwidth}} 
 \hline
 Output & Standard deviation & Top number & Equalization & Activation regularization weight & Constant regularization weight\\ \hline
 $\log y = -0.169\log^2 d+0.659\log d+4.810$ & 5 & 10 & 0 & 0.05 & 0.05\\
 \hline
\end{tabular}
\end{center}
\end{table}

Basis library: $\mathbf{\Phi} = \{+, -, \times, \div, x+c, x \cdot c, x^2, x^c, c_0x^2+c_1x+c_2\}\times 2$

\subsubsection{Wikipedia hyperlink degree distribution}\label{appendix:wiki_degree}

We fit the degree distribution of the graph of Wikipedia hyperlinks \cite{wikipedia2009} sourced from the Network Repository \cite{nr, rossi2012dynamical}. The graph consists of nodes representing individual articles in the English Wikipedia with directed edges indicating wikilinks from one article to another. We pre-process the distribution similarly to the HEP-TH citation graph, discarding nodes with degree 0 and degree less greater than 200. We fit the distribution in log-log space using the same parameters as for the HEP-TH degree distribution above.\\

\begin{table}[H]
\caption {OccamNet hyperparameters for Wikipedia hyperlink degree distribution} \label{tab:wiki}
\begin{center}
\begin{tabular}{>{\centering}p{0.2\textwidth}>{\centering}p{0.1\textwidth}>{\centering}p{0.15\textwidth}>{\centering}p{0.1\textwidth}>{\centering}p{0.18\textwidth}>{\centering\arraybackslash}p{0.18\textwidth}} 
 \hline
 Output & Standard deviation & Top number & Equalization & Activation regularization weight & Constant regularization weight\\ \hline
 $\log y = -2.586\log d+16.182$ & 5 & 10 & 0 & 0.05 & 0.05\\
 \hline
\end{tabular}
\end{center}
\end{table}

Basis library: $\mathbf{\Phi} = \{+, -, \times, \div, x+c, x \cdot c, x^2, x^c,  c_0x^2+c_1x+c_2\}\times 2$

\subsection{Cobb-Douglas production}\label{appendix:cobb}

The dataset used by Charles Cobb and Paul Douglas to empirically test the Cobb-Douglas production function contains annual capital (K), labor (L), and output (Y) data for the US manufacturing sector from 1899-1922 \cite{cobb_theory_1928}. We sweep across several values of the activation regularization weight to explore the complexity-accuracy trade-off of this dataset.

\begin{table}[H]
\caption {OccamNet hyperparameters for US manufacturing data with default bases} \label{tab:cobb_default}
\begin{center}
\begin{tabular}{>{\centering}p{0.3\textwidth}>{\centering}p{0.1\textwidth}>{\centering}p{0.15\textwidth}>{\centering}p{0.1\textwidth}>{\centering\arraybackslash}p{0.18\textwidth}} 
 \hline
 Output & Standard deviation & Top number & Equalization & Activation regularization weight \\ \hline
 $Y=\frac{26.462K}{L+22.399} + 1.129(L-30.227)$ & 50 & 10 & 5 & 0\\
 $Y=0.136K+L-11.736$ & 50 & 10 & 5 & 0.1\\
 $Y=0.0946K + L$ & 50 & 10 & 5 & 0.2\\
 $Y=1.148L$ & 50 & 10 & 5 & 0.3\\
 \hline
\end{tabular}
\end{center}
\end{table}

Basis library: $\mathbf{\Phi} = \{+, -, \times, \div, x+c, x \cdot c\}\times 3$\\

We then restrict the bases to the bare-minimum needed to construct the Cobb-Douglas function.

\begin{table}[H]
\caption {OccamNet hyperparameters for US manufacturing data with restricted bases} \label{tab:cobb_small}
\begin{center}
\begin{tabular}{>{\centering}p{0.3\textwidth}>{\centering}p{0.1\textwidth}>{\centering}p{0.15\textwidth}>{\centering}p{0.1\textwidth}>{\centering\arraybackslash}p{0.18\textwidth}} 
 \hline
 Output & Standard deviation & Top number & Equalization & Activation regularization weight \\ \hline
 $Y=K^{0.249}L^{0.754}$ & 5 & 5 & 5 & 0\\
 \hline
\end{tabular}
\end{center}
\end{table}

Restricted basis library: $\mathbf{\Phi} = \{\times, x \cdot c, x^c\} \times 3$

\subsection{Predator-prey dynamics}\label{appendix:predator_prey}

\subsubsection{Synthetic Lotka-Volterra model}\label{appendix:synth_lv}
We simulate a system of Lotka-Volterra equations as defined in \eqref{eq:lotka-volterra} with synthetic parameters and populations $H$ and $L$. The simulation spans 200 time steps with initial conditions $H_0 = L_0 = 20$ such that it contains 2 periods of the predator-prey cycle. The target ODE system is defined to be
\begin{align}
    \frac{dH}{dt} &= 0.03H-0.001HL\\
    \frac{dL}{dt} &= 0.006HL-0.15L
\end{align}
We then separately fit $\frac{dH}{dt}$ and $\frac{dL}{dt}$ since the system is decoupled.

\begin{table}[H]
\caption {OccamNet hyperparameters for synthetic Lotka-Volterra model} \label{tab:synth_lv}
\begin{center}
\begin{tabular}{>{\centering}p{0.3\textwidth}>{\centering}p{0.1\textwidth}>{\centering}p{0.15\textwidth}>{\centering\arraybackslash}p{0.18\textwidth}} 
 \hline
 Output & Standard deviation & Top number & Equalization \\ \hline
 $\frac{dH}{dt} = 0.0301H - 0.001HL$ & 5 & 1 & 5 \\
 $\frac{dL}{dt} =0.00603HL - 0.151 L$ & 5 & 1 & 5\\
 \hline
\end{tabular}
\end{center}
\end{table}

Basis library: $\mathbf{\Phi} = \{+, -, \times, \div, x+c, x \cdot c\}\times 3$

\subsubsection{Lynx-hare population}\label{appendix:lynx_hare}
One of the first datasets supporting the Lotka-Volterra model was the Hudson Bay Company's data on the lynx and hare population in Canada from 1845-1935 \cite{maclulich_fluctuations_1937}. The dataset is compiled from annual fur records on lynx and hare across 10 regions of boreal Canada, with the entries prior to 1903 being derived from questionnaires \cite{stenseth_population_1997}. Specifically, we use the subset of records from 1900-1920. Given the sparsity of the data, we utilize cubic spline interpolation to generate 100 samples at equidistant time-steps spanning the 20-year range. This allows us to compute a smoother time-series when numerically differentiating the data.

\begin{table}[H]
\caption {OccamNet hyperparameters for Lynx-Hare population data} \label{tab:lynx_hare} 
\begin{center}
\begin{tabular}{>{\centering}p{0.3\textwidth}>{\centering}p{0.1\textwidth}>{\centering}p{0.15\textwidth}>{\centering\arraybackslash}p{0.18\textwidth}} 
 \hline
 Output & Standard deviation & Top number & Equalization \\ \hline
 $\frac{dH}{dt} = 0.130H -0.00549HL-0.449$ & 5 & 1 & 0 \\
 $\frac{dL}{dt} = 0.00509HL - 0.187L + 0.188$ & 5 & 10 & 1\\
 \hline
\end{tabular}
\end{center}
\end{table}

Basis library: $\mathbf{\Phi} = \{+, -, \times, \div, x+c, x \cdot c\}\times 3$

\subsection{Compartmental models in epidemiology}\label{appendix:epidemiology}

\subsubsection{Synthetic SIR model}\label{appendix:synth_sir}
We simulate a synthetic time-series using the SIR model with initial conditions of $s_0 = 0.999, i_0=0.01, r_0 = 0$ for 60 time steps. The target model is defined to be
\begin{align}
    \frac{ds}{dt} &= -0.5si\\
    \frac{di}{dt} &= 0.5si-0.2i\\
    \frac{dr}{dt} &= 0.2i
\end{align}
We then separately fit $\frac{ds}{dt}$,  $\frac{di}{dt}$, and $\frac{dr}{dt}$.

\begin{table}[H]
\caption {OccamNet hyperparameters for synthetic SIR model} \label{tab:synth_sir} 
\begin{center}
\begin{tabular}{>{\centering}p{0.25\textwidth}>{\centering}p{0.1\textwidth}>{\centering}p{0.1\textwidth}>{\centering\arraybackslash}p{0.1\textwidth}} 
 \hline
 Output &  Standard deviation & Top number & Equalization \\ \hline
 $\frac{ds}{dt} = -0.501744si$  & 5 & 5 & 5 \\
 $\frac{di}{dt} = 0.514663si-0.209468i$  & 5 & 5 & 0 \\
 $\frac{dr}{dt} = 0.200869$  & 5 & 10 & 5 \\
 \hline
\end{tabular}
\end{center}
\end{table}

Basis library: $\mathbf{\Phi} = \{+, -, \times, \div, x+c, x \cdot c\}\times 3$

\subsubsection{Measles spread}\label{appendix:measles}
We apply OccamNet's ensemble learning capability to a panel dataset for measles infection data in the UK from 1948-1967. The measles cycle is mostly biennial, which allows us to split the data into 2-year cycles for ensemble fitting. Since there is no data on susceptible individuals, we follow the methodology outlined by Horrocks and Bauch \cite{horrocks_algorithmic_2020} to generate synthetic data for $s(t)$. We also replicate their data pre-processing by applying a Savitzky-Golay filter with a window length of 19 and polynomial order of 3 to smooth out $i(t)$. As an additional pre-processing step, we scale the derivative $\frac{di}{dt}$ by a multiplicative factor of $100$ in order to improve OccamNet loss convergence on small regression outputs.

\begin{table}[H]
\caption {OccamNet hyperparameters for measles data} \label{tab:measles} 
\begin{center}
\begin{tabular}{>{\centering}p{0.1\textwidth}>{\centering}p{0.25\textwidth}>{\centering}p{0.1\textwidth}>{\centering}p{0.1\textwidth}>{\centering\arraybackslash}p{0.1\textwidth}} 
 \hline
 Dataset & Output &  Standard deviation & Top number & Equalization \\ \hline
 1952-1953 & $\frac{di}{dt} = \frac{1}{100}(-0.282+3s+3i)$  & 5 & 5 & 1 \\
 1954-1955 & $\frac{di}{dt} = \frac{1}{100}(0.197+2s+4i)$  & 5 & 5 & 5 \\
 1956-1957 & $\frac{di}{dt} =\frac{1}{100}(-0.257+2.664s-i^2)$  & 5 & 1 & 1 \\
 1958-1959 & $\frac{di}{dt} =\frac{1}{100}(-0.28+3s+si)$  & 5 & 1 & 1 \\
 1960-1961 & $\frac{di}{dt} =\frac{1}{100}(-0.216+2s+2s^2)$  & 5 & 1 & 5 \\
 Ensemble & $\frac{di}{dt} =\frac{1}{100}(2c_0si-2i-i^2+c_0si^2)$  & 5 & 5 & 1 \\
 \hline
\end{tabular}
\end{center}
\end{table}

Basis library: $\mathbf{\Phi} = \{+, -, \times, \div, x+c, x \cdot c\}\times 3$

\subsection{Economic growth}\label{appendix:economic_growth}

\subsubsection{Synthetic Solow-Swan model}\label{appendix:synth_solow}
Twenty synthetic ``country" datasets are generated by simulating the following equations for 30 time steps:
\begin{align}
    \frac{dk}{dt} &= sy-k(\delta+g+n)\\
    \frac{dy}{dt} &= \frac{\alpha y}{k}\frac{dk}{dt}\\
    \frac{ds}{dt} &= 0.05\\
    \frac{dn}{dt} &= 0.05n
\end{align}
For each of the 20 datasets, we randomly sample the constant parameters $\alpha$, $\delta$, and $g$ as well as the initial conditions for the ODE system from the uniform distributions $\alpha \in [0.01, 0.5]$, $\delta \in [0.05, 0.1]$, $g \in [0.05, 0.1]$, $k_0 \in [0.01, 0.1]$, $s_0 \in [0.2, 0.8]$, and $n_0 \in [0.01, 0.05]$. The initial conditions for output $y_0$ are computed using the Cobb-Douglas-based assumption $y_0 = k_0^\alpha$. We use OccamNet to fit $\frac{dk}{dt}$ which is computed using the centered derivative.

\begin{table}[H]
\caption {OccamNet hyperparameters for synthetic Solow-Swan model} \label{tab:synth_solow} 
\begin{center}
\begin{tabular}{>{\centering}p{0.25\textwidth}>{\centering}p{0.1\textwidth}>{\centering}p{0.1\textwidth}>{\centering\arraybackslash}p{0.1\textwidth}} 
 \hline
 Output &  Standard deviation & Top number & Equalization \\ \hline
 $\frac{dk}{dt} = sy-k(n+c)$  & 0.5 & 5 & 1 \\
 \hline
\end{tabular}
\end{center}
\end{table}

Basis library: $\mathbf{\Phi} = \{+, -, \times, \div, x+c, x \cdot c, x^2, x^c, \log x, \exp x, \sin x, \cos x\}\times 3$\\

Unit checker: $\mathbf{U} = \begin{bmatrix} 
1 & -1\\
1 & -1\\
0 & 0\\
0 & 0
\end{bmatrix} 
$ for units [USD (millions), capita]

\subsubsection{OECD country capital growth}\label{appendix:real_solow}

We use macroeconomic data compiled by Khoo et al. \cite{khoo_neural_2021} to attempt to recover the Solow model of economic growth. Eighteen of the original 20 OECD countries are present in this dataset: Austria, Belgium, Canada, Denmark, France, Germany, Greece, Ireland, Italy, Netherlands, Norway, Portugal, Spain, Sweden, Switzerland, Turkey, United Kingdom, and United States. The time series $k(t)$, $y(t)$, $s(t)$, $n(t)$ for each country have varying starting points as well as varying lengths ranging from 4 years to 39 years.
    
\begin{table}[H]
\caption {OccamNet hyperparameters for OECD macroeconomic data} \label{tab:real_solow} 
\begin{center}
\begin{tabular}{>{\centering}p{0.2\textwidth}>{\centering}p{0.1\textwidth}>{\centering}p{0.15\textwidth}>{\centering}p{0.1\textwidth}>{\centering}p{0.18\textwidth}>{\centering\arraybackslash}p{0.18\textwidth}} 
 \hline
 Output & Standard deviation & Top number & Equalization & Activation regularization weight & Constant regularization weight \\ \hline
 $\frac{dk}{dt} = c_0\left(\frac{k^2}{y}+k\right)+c_1$  & 0.1 & 1 & 0 & 0 & 0\\
 $\frac{dk}{dt} = sy-kn$ & 0.1 & 1 & 0 & 0 & 10\\
 $\frac{dk}{dt} = sy$  & 0.1 & 1 & 0 & 1 & 10\\
 \hline
\end{tabular}
\end{center}
\end{table}

Basis library: $\mathbf{\Phi} = \{+, -, \times, \div, x+c, x \cdot c\}\times 3$\\

Unit checker: $\mathbf{U} = \begin{bmatrix} 
1 & -1\\
1 & -1\\
0 & 0\\
0 & 0
\end{bmatrix} 
$ for units [USD (millions), capita]

\end{document}